\documentclass[
aps,
prr,
reprint,
groupedaddress,
]{revtex4-2}

\usepackage{amsmath}
\usepackage{amsfonts}
\usepackage{graphicx}
\usepackage{amssymb}
\usepackage{textcomp}
\usepackage{bm}
\usepackage{color}
\usepackage{dsfont}

\usepackage{pgfplots}

\usepackage{siunitx}

\usepackage{mathtools}

\usetikzlibrary{arrows,shapes,backgrounds,calc,
    positioning,
    intersections}

\definecolor{Blue}{cmyk}{1,0.6,0,0.05}
\definecolor{Red}{cmyk}{0.04,0.87,0.89,0}
\definecolor{Green}{cmyk}{1,0.,1,0}
\definecolor{Yellow}{cmyk}{0,0.2,1,0}
\definecolor{Orange}{cmyk}{0.0,0.69,1,0}
\definecolor{Orange2}{cmyk}{0.0,0.6,1,0.00}
\definecolor{SoftCyan}{cmyk}{0.48,0.,0.07,0.11}

\definecolor{BlenderOrange}{rgb}{1, 0.5, 0.168}
\definecolor{BlenderOrangeDark}{rgb}{1., 0.43, 0.27} 

\definecolor{LaserBlue}{rgb}{0.4,0,1}
\definecolor{LaserMOT}{rgb}{1,0.372,0}
\definecolor{LaserJseven}{rgb}{1,0,0}
\definecolor{LaserJeight}{rgb}{0.35,0,0}

\pgfplotsset{colormap={CM}{color=(white) color=(Blue!50!white) color=(Blue)  color=(Blue!75!black) color=(Blue!50!black)  color=(Blue!25!black) color=(black)}}

\pgfplotsset{colormap={RedToBlue}{color=(Red)  color=(white!50!black) color=(Blue)}}

\pgfplotsset{colormap={CM2}{color=(white) color=(Blue!33!white) color=(Blue!67!white) color=(Blue)  color=(Blue!75!black) color=(Blue!50!black)  color=(Blue!25!black) color=(black)}}
\pgfplotsset{colormap={CM3}{color=(white)  color=(Blue)  color=(black)}}
\pgfplotsset{colormap={CM4}{color=(white)  color=(Red)  color=(black)}}

\DeclareMathOperator{\acos}{acos}

\newcommand{\fig}[1]{Fig.\,\ref{#1}}

\newcommand{\eqP}[1]{(\ref{#1})}

\newcommand{\bitem}{\begin{itemize}}
\newcommand{\eitem}{\end{itemize}}

\newcommand{\bti}{\begin{tikzpicture}}
\newcommand{\eti}{\end{tikzpicture}}

\newcommand{\ket}[1]{\left| #1 \right>} 
\newcommand{\bra}[1]{\left< #1 \right|} 

\newcommand{\dd}{\text{d}}

\newcommand{\hc}{\text{hc}}
\newcommand{\I}{\text{i}}

\newcommand{\E}{\text{e}}

\newcommand{\unitm}{\hat{\textbf{m}}}
\newcommand{\unitn}{\hat{\textbf{n}}}

\newcommand{\unitx}{\hat{\textbf{x}}}
\newcommand{\unity}{\hat{\textbf{y}}}
\newcommand{\unitz}{\hat{\textbf{z}}}

\newcommand{\unitmu}{\hat{\boldsymbol{\mu}}}
\newcommand{\unitnu}{\hat{\boldsymbol{\nu}}}
\newcommand{\unitxi}{\hat{\boldsymbol{\xi}}}

\newcommand{\bbold}{\textbf{b}}

\newcommand{\fbold}{\textbf{f}}

\newcommand{\pbold}{\textbf{p}}

\newcommand{\rbold}{\textbf{r}}
\newcommand{\ubold}{\textbf{u}}
\newcommand{\vbold}{\textbf{v}}
\newcommand{\wbold}{\textbf{w}}

\newcommand{\Abold}{\textbf{A}}
\newcommand{\Bbold}{\textbf{B}}

\newcommand{\Kbold}{\textbf{K}}
\newcommand{\Mbold}{\textbf{M}}

\newcommand{\Omegabold}{\boldsymbol\Omega}

\newcommand{\bas}{\begin{align}}
\newcommand{\eas}{\end{align}}

\newcommand{\bc}{\begin{center}}
\newcommand{\ec}{\end{center}}

\pgfplotsset{every axis/.append style={
  x tick label style={font=\small,yshift=0.5mm},
  y tick label style={font=\small,xshift=0.5mm},
  label style={font=\small},
  xlabel style={yshift=1mm},
  ylabel style={yshift=-3mm},
}
}

\tikzset{>=stealth}

\newcommand{\vrec}{v_{\text{rec}}}
\newcommand{\Erec}{E_{\text{rec}}}

\newcommand{\Mod}[1]{\ (\mathrm{mod}\ #1)}

\definecolor{col0}{HTML}{520BDE}
\definecolor{col1}{HTML}{F51400}
\definecolor{col2}{HTML}{E0A90B}
\definecolor{col3}{HTML}{26FF00}

\colorlet{colLaser1a}{Blue!80!black}
\colorlet{colLaser1b}{Blue!60!white}
\colorlet{colLaser2a}{Red!80!black}
\colorlet{colLaser2b}{Red!60!white}

\colorlet{col0}{Blue}
\colorlet{col1}{Red}
\colorlet{col2}{Green}

\colorlet{PRAcolor1}{Blue!100!white}
\colorlet{PRAcolor2}{Red!80!white}
\colorlet{PRAcolor3}{Green!50!white}

\newlength{\OneColumnPRLWidth}
\newcommand{\phia}{\varphi_a}
\newcommand{\phib}{\varphi_b}

\usepackage{bold-extra}

\DeclareSIUnit\gauss{G}

\newcommand{\Pabs}[1]{P_{\text{abs}}}

\def\a{x}
\def\b{z}
\def\phia{\phi_\a}
\def\phib{\phi_\b}
\def\ta{t_\a}
\def\tb{t_\b}

\begin{document}

 \title{
Realization of an atomic quantum Hall system in four dimensions
 }

\author{Jean-Baptiste Bouhiron}
\author{Aur\'elien Fabre}
\author{Qi Liu}
\author{Quentin Redon}
\author{Nehal Mittal}
\author{Tanish Satoor}
\author{Raphael Lopes}
\author{Sylvain Nascimbene}
\email{sylvain.nascimbene@lkb.ens.fr}
 \affiliation{Laboratoire Kastler Brossel,  Coll\`ege de France, CNRS, ENS-PSL University, Sorbonne Universit\'e, 11 Place Marcelin Berthelot, 75005 Paris, France}
 \date{\today}

  \begin{abstract}
 Modern condensed matter physics relies on the concept of topology to classify matter, from quantum Hall systems to topological insulators. Engineered systems, benefiting from synthetic dimensions, can potentially give access to novel topological states predicted in dimensions $D > 3$. We report the realization of an atomic quantum Hall system evolving in four dimensions (4D), with two spatial dimensions and two synthetic ones encoded in the large spin of dysprosium atoms. The non-trivial topology is evidenced by  measuring a quantized electromagnetic non-linear response and observing anisotropic hyperedge modes. We also excite non-planar cyclotron motion, contrasting with its circular equivalents in $D\leq3$. 
Our work opens to the investigation of strongly-correlated topological liquids in 4D generalizing fractional quantum Hall states.
 \end{abstract}
 
 \maketitle

Topological order plays a central role in the classification of states of matter beyond Landau's paradigm of symmetry breaking. It was originally introduced to explain the Hall conductance of two-dimensional (2D) quantum Hall systems, quantized by the first Chern number \cite{klitzing_new_1980,thouless_quantized_1982}. Since then, various forms of topological systems, such as  3D topological insulators \cite{hsieh_topological_2008} and Weyl semi-metals \cite{lu_experimental_2015,xu_discovery_2015}, have been explored  in condensed matter \cite{bansil_colloquium_2016}. A larger variety of topological systems have been predicted, and organized in a ten-fold-way classification scheme according to the system symmetries and dimensionality \cite{ryu_topological_2010}. In particular, higher-dimensional systems, which can be explored in engineered systems with synthetic dimensions \cite{celi_synthetic_2014,ozawa_topological_2019}, can host special classes of topological matter \cite{ryu_topological_2010}, in particular a generalization of the quantum Hall effect in 4D \cite{frohlich_new_2000,zhang_four-dimensional_2001,karabali_quantum_2002}. Different protocols have been proposed \cite{kraus_four-dimensional_2013,price_four-dimensional_2015,ozawa_synthetic_2016} to realize 4D Hall insulators, both for time-reversal invariant systems (classes AI and AII) and in the absence of discrete symmetry (class A).  In those systems, the non-trivial topology leads to specific behaviour, such as the quantization of  the non-linear response to both electric and magnetic perturbations, characterized by the second Chern number $\mathcal{C}_2$ \cite{zhang_four-dimensional_2001,qi_topological_2008,price_four-dimensional_2015} -- a topological invariant also relevant for tensor monopoles in high dimensions \cite{yang_generalization_1978,sugawa_second_2018,tan_experimental_2021,chen_synthetic_2022}. The topology of a 4D quantum Hall insulator (class A) also gives rise to  anisotropic motion close to a 3D hyperedge of the system, ballistic along a given orientation, and still prohibited along the two remaining directions of the hyperedge \cite{elvang_quantum_2003,karabali_effective_2004}.

\begin{figure}[!t]
 \bc
 \includegraphics[
 draft=false,scale=0.8,
 trim={3mm 2mm 0 0.cm},
 ]{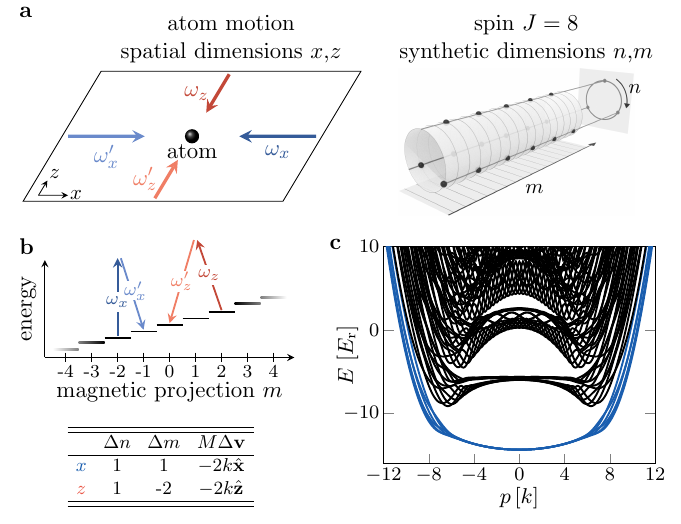}
 \ec
\caption{
\textbf{Scheme of the 4D atomic system.} (\textbf{A}) The atomic motion in the $xz$ plane is coupled to the internal spin $J=8$ using two-photon optical transitions along $x$ and $z$ (blue and red arrows, respectively). The spin encodes two synthetic dimensions given by the magnetic projection $m$ and its remainder $n=m\Mod{3}$ of its Euclidian division by 3, leading to a synthetic space of cylindrical geometry \cite{fabre_simulating_2022}. (\textbf{B}) Scheme of the light-induced spin transitions, of first- and second-order along $x$ and $z$, respectively. They induce correlated spin-orbit dynamics, with distinct hopping along $n$ and $m$ according to the rules given in the table.   (\textbf{C}) Dispersion relation plotted as a function of the momentum $p$, for 6 values of the quasi-momentum $q$ uniformly spanning the Brillouin zone. The ground band is pictured as blue lines.
\label{fig_scheme}}
\end{figure}

So far, topological properties linked with the 4D Hall effect have been revealed via geometrical charge pump experiments in 2D systems \cite{lohse_exploring_2018,zilberberg_photonic_2018}. A truly 4D Hall system  has also been realized using electronic circuits -- however, no direct evidence of topological quantization has been reported \cite{wang_circuit_2020}. Here, we engineer an atomic quantum Hall system evolving in 4D, by coupling with light fields two spatial dimensions $x$ and $z$ and two synthetic ones encoded in the electronic spin $J=8$ of dysprosium atoms \cite{celi_synthetic_2014,mancini_observation_2015,stuhl_visualizing_2015} (\fig{fig_scheme}A). As a signature of non-trivial topology, we measure a quantized electromagnetic non-linear response and observe anisotropic  hyperedge modes. We also probe low-lying excitations, revealing complex cyclotron orbits beyond the planar circular paradigm occurring in lower dimensions $D\leq3$.

The coupling between motion and spin degrees of freedom is generated by a pair of lasers counter-propagating along $x$ (resp. $z$) and resonantly driving spin transitions $m\rightarrow m+1$ (resp. $m\rightarrow m-2$), while imparting a momentum kick $-2k\unitx$ (resp. $-2k\unitz$) \cite{lin_synthetic_2009} (\fig{fig_scheme}B). Here, $m$ is the spin projection along $z$ ($-J\leq m\leq J$, $m$ integer), $k=2\pi/\lambda$ is the light momentum for a wavelength $\lambda=\SI{626.1}{\nano\meter}$, and we assume a unit reduced Planck constant $\hbar=1$.
The atom dynamics is described by the Hamiltonian
\begin{align}
H&=\frac{Mv^2}{2}-\bigg(\ta\E^{\I \phia}\frac{J_+}{J}+\tb\E^{\I \phib}\frac{J_-^2}{J^2}+\hc\bigg)+\beta\frac{J_z^2}{J^2},\label{eq_H}
\end{align}
where $\vbold$ is the atom velocity and $\phi_\alpha=-2k\alpha$ is the relative phase of the two laser beams involved in each Raman process $\alpha=x,z$. The laser intensities and polarisations control the amplitudes $t_\alpha$ and the  quadratic Zeeman shift  $\beta=-2\tb$ \cite{supp_mat}.

The laser-induced spin transitions can be interpreted as hopping processes  in a two-dimensional synthetic space $(m,n)$ involving the spin projection $m$ and the remainder $n=m\Mod{3}$ of its Euclidian division by 3 (with $n=0,1,2$) \cite{fabre_simulating_2022}. While the first-order spin coupling $J_+$ acts on these two dimensions in a similar manner (hopping $\Delta n^{(x)}=\Delta m^{(x)}=1$), the second-order coupling $J_-^2$ induces hoppings $\Delta n^{(z)}=1$ and $\Delta m^{(z)}=-2$ leading to differential dynamics along $m$ and $n$.
The complex phases  $\phi_\alpha$ (with $\alpha=x,z$) can be interpreted as Peierls phases upon the hopping of a charged particle on a lattice subjected to a magnetic field.
Assuming unit charge, we write $\phi_\alpha=\int A_\beta\dd r^\beta=A_n\Delta n^{(\alpha)}+A_m\Delta m^{(\alpha)}$, leading to the explicit expression for the vector potential 
\begin{equation}
 \Abold=\frac{1}{3}(0,0,2\phi_x+\phi_z,\phi_x-\phi_z)_{x,z,n,m}.\label{eq_A}
\end{equation}
The magnetic field is then defined by the anti-symmetric tensor $B_{\alpha\beta}=\partial_\alpha A_\beta-\partial_\beta A_\alpha$, as
 \begin{equation}
  \Bbold=\frac{2k}{3}\begin{pmatrix}
     0 & 0 & -2 & -1\\
     0 & 0 & -1 & 1\\
     2 & 1 & 0 & 0\\
     1 & -1 & 0 & 0
    \end{pmatrix}.\label{eq_Bfield}
\end{equation}
Similarly to the 2D quantum Hall effect, this magnetic field gives rise to an energy separation between  quasi-flat magnetic Bloch bands. Within each band, motion becomes effectively two-dimensional, with the guiding center coordinate along $n$ canonically conjugated to the position along $\unitnu=(2\unitx+\unitz)/\sqrt5$, while $m$ is conjugated to the projection on $\unitmu=(\unitx-\unitz)/\sqrt2$. The energy levels are indexed by the canonical momentum $\pbold=M\vbold+2k m\unitx\Mod{\Kbold}$, which  is conserved in the absence of external force. Here, the reciprocal lattice vector $\Kbold=2k(2\unitx+\unitz)\parallel \unitnu$ corresponds to the momentum kick imparted on a non-trivial cycle \mbox{$m\overset{z}{\rightarrow} m+2\overset{x}{\rightarrow} m+1\overset{x}{\rightarrow} m$} involving one transition along $z$ and two along $x$. In the following, we decompose the momentum as $\pbold=p\unitxi+q\unitnu$, with  $\unitxi=(\unitx-2\unitz)/\sqrt5\perp\unitnu$, such that the first Brillouin zone is defined for $|q|<K/2$ and arbitrary $p$. The energy levels of the Hamiltonian \eqP{eq_H} organize in Bloch bands shown in \fig{fig_scheme}d. We focus here on the ground band, which is quasi-flat in the bulk mode region $|p|\lesssim7k$ \cite{supp_mat}.

Our experiments use ultracold dilute samples of $\simeq3.0(3)\times10^4$ atoms of $^{162}$Dy, prepared in an optical dipole trap at a  temperature $T=\SI{260(10)}{\nano\kelvin}$. The atoms are subjected to a magnetic field $B=\SI{221(1)}{\milli G}$ along $z$, and initially spin-polarized in the magnetic sub-level $m=-J$. We adiabatically ramp up the laser intensities to generate the spin couplings described in \eqP{eq_H} with 
$t_x=5.69(6)\,\Erec$ and $t_z=5.1(1)\,\Erec$, where $\Erec=k^2/(2M)$ is the recoil energy. Starting in the $m=-J$ edge mode region with $p<7k$, we prepare  arbitrary momentum states  of the ground energy band by applying a weak force on a typical $\SI{1}{\milli\second}$ timescale \cite{supp_mat}.  At the end of the experiments, we probe the velocity distribution by imaging the atomic sample after  free expansion in the presence of a magnetic field gradient, such that the different $m$ states are spatially separated.

\begin{figure}[!t]
\bc
 \includegraphics[
 draft=false,scale=0.9,
 trim={1.5mm 2mm 0 0.cm},
 ]{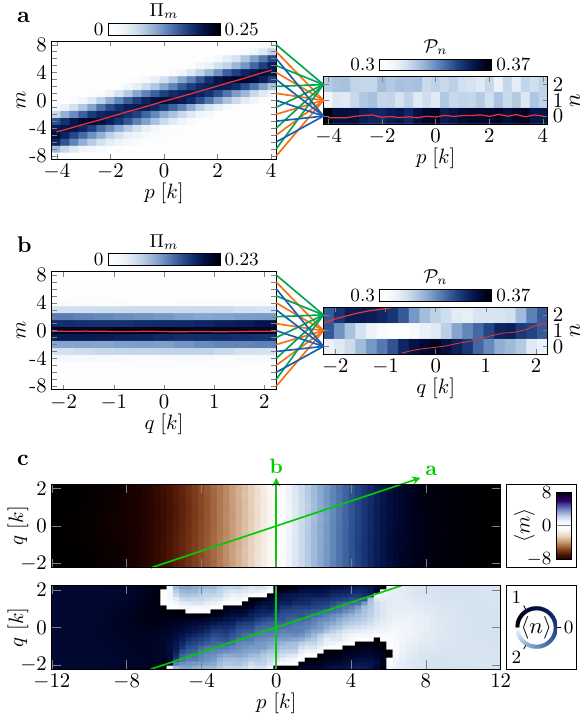}
 \vspace{-8mm}
 \ec
\caption{
\textbf{Hall drift along the synthetic dimensions.} (\textbf{A}) Evolution of the measured spin projections $\Pi_m$ and $\mathcal{P}_n$ as a function of $p$ upon adiabatic driving along $\unitmu$ -- the spatial direction conjugated with $m$. The mean values $\langle m\rangle$ and $\langle n\rangle$ (computed as $\tfrac{3}{2\pi}\arg\langle\E^{\I 2\pi m/3}\rangle)$ are shown as red lines. (\textbf{B}) Same quantities plotted for a  driving along $\unitnu$ -- the spatial direction conjugated with $n$. (\textbf{C}) Measurements of the mean values $\langle m\rangle$ and $\langle n\rangle$ in the Brillouin zone. The green arrows represent the driving directions considered in \textbf{A} and \textbf{B}. 
\label{fig_spin_pumping}}
\end{figure}

\begin{figure*}[t!]
 \includegraphics[
 draft=false,scale=0.885,
 trim={1.5mm 4.mm 0 1mm},
 ]{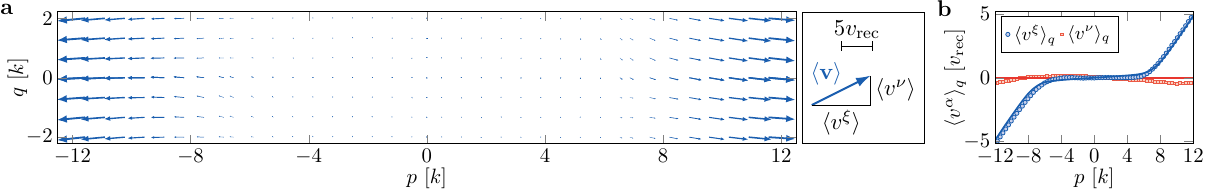}
\caption{
\textbf{Frustration of motion in the bulk and anisotropic ballistic edge modes.} (\textbf{A}) Evolution of the mean velocity $\langle \vbold\rangle$ versus momentum. The arrow is scaled according the mean velocity modulus. (\textbf{B}) Measurements of the $q$-average velocity components versus $p$. The solid lines are the expected variations for the ground band of the Hamiltonian \eqP{eq_H}.
\label{fig_velocity}}
\end{figure*}

We first investigate the anomalous Hall drift in spin space  upon the application of  a weak force in the $xz$ plane. For a force oriented along $\unitmu$ (spatial direction conjugated to $m$), the  spin projection probabilities $\Pi_m$ reveal a drift of the mean spin projection $\langle m\rangle$, while the mean remainder $\langle n\rangle$ remain approximately constant (\fig{fig_spin_pumping}a). An opposite behavior is observed when applying a force along  $\unitnu$ (direction conjugated to $n$), with a quasi-linear variation of $\langle n\rangle$ while  $\langle m\rangle$ remains constant (\fig{fig_spin_pumping}b). 
More generally, in the bulk of the system, where the band dispersion  can be neglected, the variation with momentum of the mean values  $\langle n\rangle$ and $\langle m\rangle$ can be expressed as an anomalous Hall drift governed by the antisymmetric Berry curvature tensor $\Omegabold_{\text{bulk}}$, as \cite{price_measurement_2016,estienne_ergodic_2021}
\begin{align}\label{eq_Hall_drift}
 \dd \langle r^\alpha\rangle&=\Omega_{\text{bulk}}^{\alpha\beta}\dd  p_\beta,\\
 \Omegabold_{\text{bulk}}&=\Bbold^{-1}=\frac{1}{2k}\begin{pmatrix}
    0 & 0 & 1 & 1\\
    0 & 0 & 1 & -2\\
    -1 & -1 & 0 & 0\\
    -1 & 2 & 0 & 0
   \end{pmatrix},\label{eq_Omega_bulk}
\end{align}
where $\rbold$ is the position vector and $\dd\pbold$ is the momentum variation  due to the external force.  
We confront this prediction to our measurements of the mean positions $\langle m\rangle$ and $\langle n\rangle$ as a function of $\pbold$ (\fig{fig_spin_pumping}c). In the center of the Brillouin zone  $|p|\leq 4k$ (bulk mode region), we fit the measured Hall drift with the linear function \eqP{eq_Hall_drift}, yielding  $\{\Omega^{nx},\Omega^{nz},\Omega^{mx},\Omega^{mz}\}=\{-1.00(2),-0.98(2),-0.98(2),1.96(2)\}/(2k)$, consistent with \eqP{eq_Omega_bulk}. We do not measure any significant spatial drift upon the application of a force in the $xz$ plane, consistent with $\Omega^{xz}=0$. Similarly, no Hall current along $n$ is measured when applying a  force along $m$ (corresponding to a perturbative Zeeman field), compatible with $\Omega^{nm}=0$ \cite{supp_mat}.

Further insight on the ground band properties is provided by the mean velocity $\langle\vbold\rangle$ (\fig{fig_velocity}), which reveals distinct behaviours between bulk and the edge modes. 

In the bulk $|p|\lesssim7k$, the mean velocity remains much smaller than the recoil velocity $\vrec=k/M$ (\fig{fig_velocity}b). Since $\langle\vbold\rangle=\nabla_{\pbold} E_0$, it confirms that the ground band energy $E_0$ is quasi-flat in the bulk (\fig{fig_scheme}d). This measurement illustrates the frustration of motion induced by the magnetic field, similarly to flat Landau levels in 2D electron gases.

For $p\gtrsim7k$, the atoms mostly occupy the edge $\rbold^{\text{edge}}= J\unitm$ of the synthetic dimension $m$. We measure a non-zero mean velocity, whose $\xi$ component increases with $p$, while the $\nu$ projection remains small (\fig{fig_velocity}b). This observation is characteristic of an anisotropic edge mode of a 4D quantum Hall system, which corresponds to a collection of 1D conduction channels oriented along the direction $\wbold_{\text{motion}}$, with
$
w_{\text{motion}}^\alpha\propto \Omega_{\text{bulk}}^{\alpha \beta}r_\beta^{\text{edge}},
$
here corresponding to the  direction $\unitxi$ \cite{elvang_quantum_2003}. Within the edge, the motion in the plane orthogonal to $\wbold_{\text{motion}}$ remains inhibited, in agreement with the measured $\nu$ velocity. A similar behaviour is found on the edge $-J\unitm$, albeit with opposite orientation of velocity.

\begin{figure}[t]
\bc
 \includegraphics[
 draft=false,scale=1,
 trim={2mm 2mm 0 0.cm},
 ]{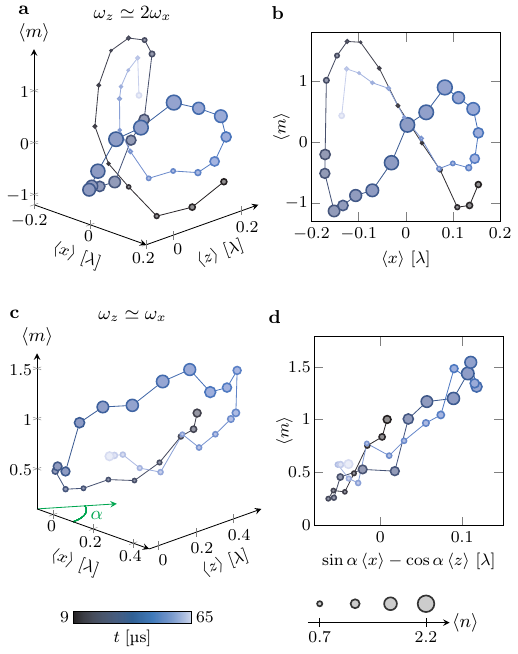}
 \ec
\caption{
\textbf{Cyclotron dynamics.} (\textbf{A}) Evolution of the center-of-mass position following a velocity kick, for $\omega_z\simeq2\omega_x$. The cartesian coordinates correspond to $(x,z,m)$, while $n$ is given by the mark size. The mark color indicates the evolution time subsequent to the kick. (\textbf{B}) Projection in the $xm$ plane corresponding to a quasi-closed Lissajouss curve. 
 (\textbf{C}) Cyclotron orbit measured for $\omega_z\simeq\omega_x$. The green arrow shows the viewpoint for the planar projection shown in \textbf{D}. (\textbf{D}) Two-dimensional projection  revealing the planar nature of the orbit viewed from the side. 
\label{fig_orbits}}
\end{figure}

A hallmark of 4D quantum Hall physics is the peculiar nature of excitations above the ground band, which can be linked to classical cyclotron trajectories. While cyclotron motion in 2D and 3D always corresponds to planar circular orbits, we expect more complex trajectories in 4D, involving  two planar rotations occurring at different rates.  In our system, each of these two elementary excitations is generated by the Raman coupling along $x$ or $z$, of  corresponding frequencies $\omega_x$ and $\omega_z$ independently set by the amplitudes $t_x$ and $t_z$. 
We excite the atoms by applying a diabatic velocity kick, and measure the subsequent time evolution of the center of mass. 
We show in  \fig{fig_orbits}a the orbit measured for $\omega_z/\omega_x\simeq 2$, revealing a non-planar trajectory. For this integer frequency ratio, the orbit is almost closed, and is reminiscent of a Lissajous orbit (\fig{fig_orbits}b). We also studied the case of degenerate frequencies $\omega_z\simeq\omega_x$, which correspond to the coupling amplitudes used for our study of the ground band. In this `isoclinic' case, we recover a planar cyclotron motion akin to lower-dimensional cyclotron orbits (\fig{fig_orbits}c,d).


The non-trivial  topology of a quantum Hall system manifests in the quantization of its  electromagnetic response. While involving the linear Hall conductance in 2D, it requires in 4D  considering the non-linear response to both an electric force $\fbold$ and a magnetic field $\bbold$. These two perturbations induce a current
\[
j^\alpha_{\text{non-linear}}=\frac{\mathcal{C}_2}{4\pi^2}\epsilon^{\alpha\beta\gamma\delta}f_\delta b_{\beta\gamma},
\]
where $\epsilon^{\alpha\beta\gamma\delta}$  is the 4D Levi-Civita symbol and $\mathcal{C}_2$ is the integer second Chern number \cite{price_four-dimensional_2015}. In other words, the magnetic field $b_{\beta\gamma}$ induces a Hall effect in the perpendicular plane.

\begin{figure*}[t!]
 \includegraphics[
 draft=false,scale=1,
 trim={1mm 2mm 0 0.cm},
 ]{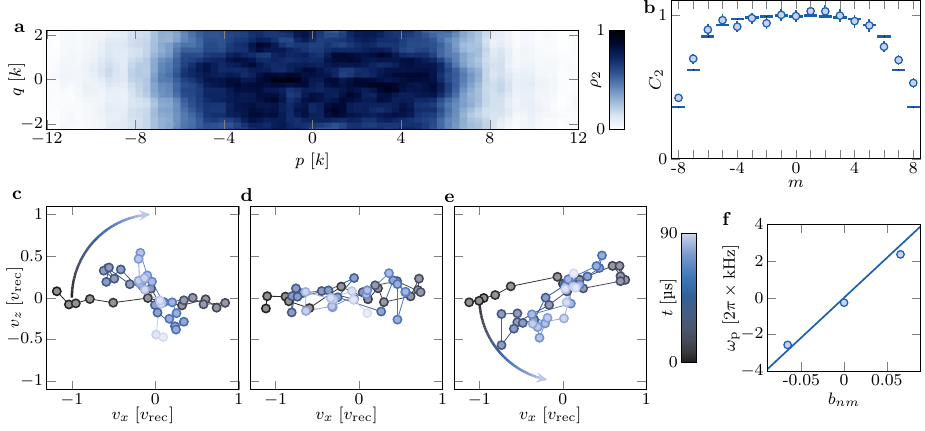}
\caption{
\textbf{Topological structure of the ground band.} (\textbf{A}) Second Chern character $\rho_2$ computed from the product of measured Berry curvatures. (\textbf{B}) Local second Chern marker inferred from \textbf{A}. (\textbf{C}), (\textbf{D}), (\textbf{E}) Velocity dynamics in the $xz$ plane for three different values of the magnetic perturbation $b_{nm}=-0.066,0,0.066$, respectively. (\textbf{F}) Precession rate $\omega_{\text p}$ as a function of the $b_{nm}$ field. The solid line is the prediction of \eqP{eq_omegap} for the bulk second Chern character $\rho_2^{\text{bulk}}=3/(4k^2)$.
\label{fig_precession}}
\end{figure*}

We first demonstrate the quantization of the non-linear response by reconstructing the second Chern number in terms of a non-linear combination of Berry curvatures. For an infinite lattice system, it can be expressed as an integral over the Brillouin zone $\mathcal{C}_2=\int\rho_2\,\dd^4 q/(8\pi^2)$,
where $\rho_2=\tfrac{1}{8}\epsilon_{\alpha\beta\gamma\delta}\Omega^{\alpha\beta}\Omega^{\gamma\delta}$ is the second Chern character (see \fig{fig_precession}a and \cite{supp_mat}). Since our system exhibits edges along $m$, the full band spectrum is gapless (\fig{fig_scheme}d), preventing a truly topological character for the system as a whole. To recover quantization, we compute from our data the local second Chern marker, which quantifies the non-linear response resolved in $m$, as \cite{supp_mat}
\[
C_2(m)=\frac{1}{3}\int_{|p|<p^\star}\rho_2(p,q)\Pi_m(p,q)\dd p\,\dd q,
\]
where $p^\star=7k$ is a momentum cutoff delimiting the bulk mode region (\fig{fig_precession}b). In the central region $-5\leq m\leq 5$, we measure an almost constant  marker $C_2(m)=0.97(6)$, compatible with the second Chern number $\mathcal{C}_2=1$, demonstrating topological quantization.

In order to directly probe the non-linear response, we implement a magnetic perturbation $b_{nm}$, expecting the appearance of a Hall effect in the $xz$ plane. For this, we modified the polarization of the laser beam propagating along $x$, such that the spin transition amplitude becomes \cite{supp_mat}
\[
J_+\rightarrow J_++\I \epsilon\frac{\{J_+,J_z\}}{2},
\] 
with $\epsilon\simeq0.1$, leading to a  complex phase amplitude $\phi_x'\simeq \epsilon m$. According to \eqP{eq_A}, this phase corresponds to an additional  vector potential $\Abold'=(0,0,2\epsilon m,\epsilon m)/3$, hence the magnetic field component  $b_{nm}=-\partial_m A_n=-2\epsilon/3$. We investigate its effect on the $xz$ dynamics subsequent to a kick along $x$, using laser couplings such that $\omega_x\simeq\omega_z\equiv\omega$ in the absence of magnetic perturbation. The $xz$ dynamics shown in \fig{fig_precession}c,d,e are reminiscent of those of a  Foucault pendulum, namely a harmonic oscillation at $\omega$, with a slower precession governed by the $b_{nm}$ field. We show in \cite{supp_mat} that, in the bulk, the precession rate 
\begin{equation}\label{eq_omegap}
\omega_{\text p}=\frac{1}{2}M\omega^2\rho_2^{\text{bulk}}b_{nm}
\end{equation}
gives access to the second Chern character $\rho_2^{\text{bulk}}$. Our measurements yield $\rho_2^{\text{bulk}}\simeq0.65(2)/k^2$, close to the theoretical expectation $\rho_2^{\text{bulk}}=\tfrac{1}{8}\epsilon_{\alpha\beta\gamma\delta}\Omega_{\text{bulk}}^{\alpha\beta}\Omega_{\text{bulk}}^{\gamma\delta}=3/(4k^2)$ and to the value $0.75(7)/k^2$ computed from the Berry curvature measurements.

Other specific properties of the 4D quantum Hall effect could be addressed in future work, such as the complex edge mode trajectories expected for compact boundaries \cite{estienne_ergodic_2021}. Our protocol could also be generalized to engineer other classes of topological systems, such as 5D Weyl semimetals \cite{lian_five-dimensional_2016} and 6D
quantum Hall systems \cite{lee_electromagnetic_2018,petrides_six-dimensional_2018}. 

Furthermore, our work paves a way towards the study of interacting quantum many-body physics in high-dimensional topological structures. Interactions between atoms should drive a redistribution of momentum across the ground band, leading to a many-body ground state involving the entire spin manifold. For bosonic atomic gases in the regime of large filling of the ground-band orbitals, we expect the formation of a Bose-condensed superfluid exhibiting quantized vortices \cite{mccanna_superfluid_2021}. For small fillings, the system should  form strongly-correlated liquids generalizing  fractional quantum Hall states \cite{zhang_four-dimensional_2001,nayak_non-abelian_2008,bernevig_effective_2002}. A prerequisite to these studies would be to enlarge further the size of the synthetic dimensions, such that localized excitations can be properly defined and manipulated.

\acknowledgements

We thank B. Estienne and N. Goldman for discussions, and J. Dalibard for insightful comments on the manuscript. Funding: This work is supported by European Union (grant TOPODY 756722 from the European Research Council) and Institut Universitaire de France. N.M. acknowledges support from DIM Quantip of r\'egion
\^Ile de France. Source data, software codes for the figures of the main text and supplementary materials, and other datasets generated and analyzed during the current study are available at https://zenodo.org/records/10606567.

%

\cleardoublepage

\onecolumngrid

\setcounter{equation}{0}
\setcounter{figure}{0}
\setcounter{table}{0}
\makeatletter
\renewcommand{\theequation}{S\arabic{equation}}
\renewcommand{\thefigure}{S\arabic{figure}}

\newcommand{\titlesupp}[1]{{\large\noindent\textbf{#1}} \\[0.25mm]}
\newcommand{\titlemethods}[1]{\noindent\textbf{#1} \\[0.5mm]}
\newcommand{\subtitlemethods}[1]{\noindent\emph{#1} \\[0.25mm]}

\begin{center}
\textbf{Supplementary materials for}\\
\bigskip
{\large\textbf{Realization of an atomic quantum Hall system in four dimensions}}\\ 
\bigskip
Jean-Baptiste Bouhiron, Aurélien Fabre, Qi Liu, Quentin Redon, \\Nehal Mittal, Tanish Satoor, Raphael Lopes, Sylvain Nascimbene\\
\medskip
Correspondence to sylvain.nascimbene@lkb.ens.fr\\
\medskip
Laboratoire Kastler Brossel, Collège de France, CNRS, ENS-PSL University, \\
Sorbonne Université, 11 Place Marcelin Berthelot, 75005 Paris, France

\end{center}

\bigskip

\twocolumngrid

\titlesupp{Materials and Methods}

All error bars are the $1-\sigma$ statistical uncertainty computed from a bootstrap random sampling of experimental data.

\bigskip
\titlemethods{Algebra of light-induced spin transitions}
The Raman beams are produced by a laser whose frequency is close to the $\SI{626.1}{\nano\meter}$ optical transition of atomic $^{162}$Dy (line width $\Gamma\simeq\SI{0.85}{\micro\second^{-1}}$). The detuning from resonance $\Delta=-2\pi\times\SI[parse-numbers = false]{7.0(1)}{\giga\hertz}$ is large enough for incoherent Rayleigh scattering to be negligible on the experiment timescale. 

For most experiments except those reported in \fig{fig_precession}c-f, (those with no magnetic perturbation $b_{nm}=0$), the $x$-Raman beams are linearly polarized along $\cos\theta\unity\pm\sin\theta\unitz$, with $\theta\simeq\acos(1/\sqrt3)$ thus cancelling the rank-2 tensor light shift $J_z^2$. Denoting $\omega_{\text Z}+\delta_x$ the frequency difference between both beams, the spin-dependent light shift produced by the $x$-Raman beams writes, within the rotating wave approximation,
\begin{align*}
V_x&=-t_x\left(\E^{\I(-2kx+\delta_x t)}\frac{J_+}{J}+\hc\right),\\
t_x&=\frac{4\sqrt2(2J+3)}{3(J+1)(2J+1)}\frac{3\pi c^2}{2\omega_0^3}\frac{\Gamma}{\Delta}I_x,
\end{align*}
where $I_x$ is the intensity of each Raman laser beam and $\omega_0$ is the optical resonance frequency. The $z$-Raman beams are circularly polarized $\sigma_\pm$, and differ in frequency by $2\omega_{\text Z}+\delta_z$, leading to the light shift
\begin{align*}
V_z&=-t_z\left(\E^{\I(-2kz+\delta_z t)}\frac{J_-^2}{J^2}+\hc\right)+\beta\frac{J_z^2}{J^2},\\
t_z&=-\frac{J^2}{2(J+1)(2J+1)}\frac{3\pi c^2}{2\omega_0^3}\frac{\Gamma}{\Delta}I_z,\\
\beta&=-2t_z,
\end{align*}
where $I_z$ is the intensity of each beam. 
The time-dependency associated to the detunings $\delta_x$ and $\delta_z$ can be suppressed by considering the system in the reference frame moving at velocity $\vbold_{\text{frame}}=(\delta_x\unitx+\delta_z\unitz)/(2k)$, leading to the static Hamiltonian given by eq.\,\eqP{eq_H}. We interpret all our experiments in this moving frame, and use time-dependent frequency detunings $\delta_x(t)$ and $\delta_z(t)$ to exert an inertial force to the atoms, as described in the next section. 

The bandstructure of the Hamiltonian \eqP{eq_H}, shown in \fig{fig_scheme}C of the main text, is computed using standard techniques, using a finite basis of momentum states with momentum cutoff $p_\star=20\hbar k$. The mean velocity of ground-band momentum states, shown in \fig{fig_velocity}B, is obtained as the derivative of the ground-band energy $\langle\vbold\rangle=\partial_{\pbold}E_0$.

The laser beam waists $w\simeq\SI[parse-numbers = false]{40}{\micro\meter}$ for each Raman beam give rise to a confinement transverse to the beam propagation axis. We measured confinement frequencies  $\omega_x^{\text{trap}}=2\pi\times\SI{78(3)}{\hertz}$ and  $\omega_z^{\text{trap}}=2\pi\times\SI{122(2)}{\hertz}$ for atoms polarized in $m=-J$. The corresponding oscillations periods being much longer than the experiment timescale of $\SI{1}{\milli\second}$, we neglect the effect of this confinement in our analysis.

The magnetic perturbation $b_{nm}$ is produced by changing the polarization of one Raman laser propagating along $x$ to a circular polarization $\sigma_\pm$, leading to a spin-dependent light shift (written in the moving frame)
\begin{align*}
V_x'&=-t_x'\E^{-\I2kx}\frac{J_+ +(\gamma+\I\epsilon)\{J_+,J_z\}/2}{J}+\hc+\beta'\frac{J_z^2}{J^2},
\end{align*}
where $t_x'=3t_x/4$, $\beta'=J^2t_x'/4\sqrt 2(2J+3)$, $\epsilon=\pm4\sqrt2/3(2J+3)$ and $\gamma=|\epsilon|/\sqrt 8$. The parameter $\epsilon$ is directly linked to the magnetic perturbation
\[
b_{nm}=-\frac{2\epsilon}{3}=\mp\frac{8\sqrt2}{9(2J+3)}\simeq\mp0.0661.
\]
We compensate the decrease of $t_x$ by increasing the intensities of the $x$-Raman beams by a factor $4/3$ compared to the case $b_{nm}=0$. The parameters $\beta'$ and $\gamma$ induce a small deformation of the band structure, without significantly modifying the non-linear transverse response of the system, which is globally topologically protected.

\begin{figure}[!t]
 \includegraphics[
 draft=false,scale=0.85,
 trim={1mm 2mm 0 0.cm},
 ]{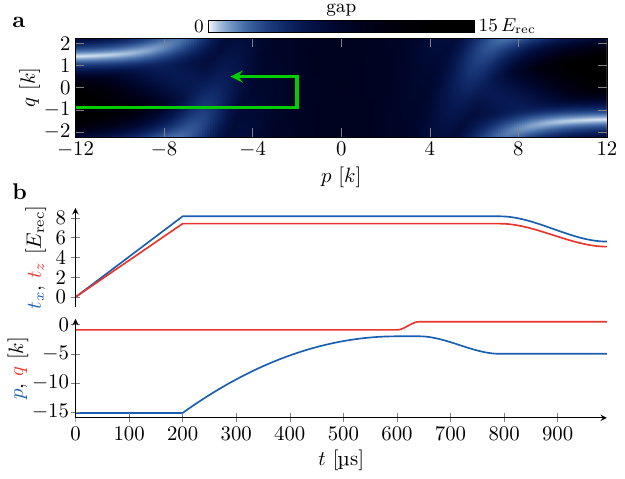}
\caption{
\textbf{Momentum state preparation.} \textbf{a.} Theoretical variation of the energy gap with momentum, and sketch of the momentum ramp used to prepare an arbitrary state (green arrow). It avoids crossing the lines of small energy gaps (in white) occurring in the edge mode region.
\textbf{b.} Time evolution of the light couplings $t_x$, $t_z$ and momentum components $p$, $q$ for the adiabatic preparation of a given momentum state. 
\label{fig_momentum}}
\end{figure}

\medskip
\titlemethods{Preparation of arbitrary momentum states} 
The atomic gas is initially prepared in a crossed optical dipole trap (oscillation frequencies $\omega_x=2\pi\times\SI{263(5)}{\hertz}$ and $\omega_z=2\pi\times\SI{140(6)}{\hertz}$), at a sub-recoil temperature $T=\SI{260(10)}{\nano\kelvin}$ using standard cooling techniques. The atoms are subjected to a magnetic field $B=\SI{221(1)}{mG}$ along $z$, producing a linear Zeeman splitting between the magnetic sub-levels $m$ of frequency $\omega_{\text Z}=2\pi\times\SI{385(3)}{\kilo\hertz}$. At this stage, the atoms are fully-polarized in the ground state $m=-J$.

The preparation of a given momentum state dressed by the Raman lasers proceeds as follows. We switch off the optical dipole trap, and increase the Raman laser intensities to $t_x=8.2(1)\,\Erec$ and $t_z=7.4(1)\,\Erec$ in $\SI{200}{\micro\second}$. During this process, the Raman lasers are off-resonant, with frequency detunings  $\delta_x=-\delta_z/2=14.9(2)\,\Erec$ corresponding to momentum components $p=-15.5(1)\,k$ and $q=-2k/\sqrt5\simeq-0.89\,k$. For this momentum, the ground state  matches the initial spin polarization in $m=-J$. We then ramp the detunings towards resonance, leading to an inertial force when considering the atoms in the moving frame as discussed above. The preparation of  given $\pbold$ states is based on momentum trajectories  as  pictured in \fig{fig_momentum}a. We first increase the momentum to $p=-2\,k$ in the bulk mode region, then ramp the quasi-momentum $q$, followed by a variation of $p$ to the final value. We finally decrease the spin coupling amplitudes to $t_x=5.69(6)\,\Erec$ and $t_z=5.1(1)\,\Erec$. This protocol minimizes the crossing of small energy gaps separating the ground and first excited bands in the edge mode region. The duration of detuning ramps, displayed in \fig{fig_momentum}b, is chosen as a compromise between the adiabaticity requirement and the  effect of dipolar relaxation, which leads to heating of the sample and atom loss on a $\SI{10}{\milli\second}$ timescale.

\begin{figure}[!t]
 \includegraphics[
 draft=false,scale=0.82,
 trim={2mm 2mm 0 0.cm},
 ]{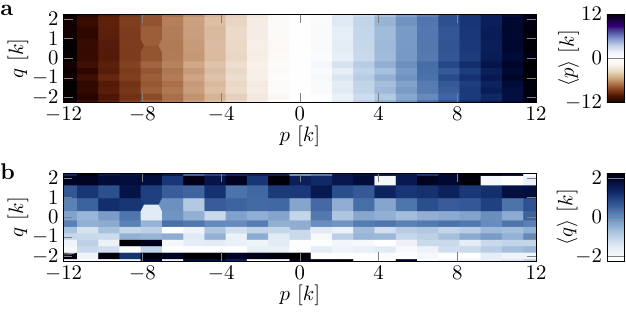}
\caption{
\textbf{Momentum projections.} Measurement of the mean momentum components $\langle p\rangle$ (\textbf{a}) and $\langle q\rangle$ (\textbf{b}) of a thermal gas as a function of the targeted momentum.  
\label{fig_momentum_fits}}
\end{figure}

We show in \fig{fig_momentum_fits} a comparison between measured and targeted momenta across the Brillouin zone. The small discrepancy between theory and experiment can be attributed to the confinement induced by the Raman lasers and gravity, which provide an additional small contribution to the momentum dynamics. This effect does not play a role in our data analysis, as explained in the next section.

\begin{figure}[!t]
 \includegraphics[
 draft=false,scale=0.92,
 trim={3mm 2mm 0 0.cm},
 ]{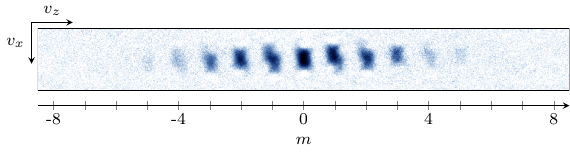}
\caption{
\textbf{Typical absorption image.} Absorption image of an atomic sample of mean momentum $\langle\pbold\rangle\simeq 0$ and thermal width $\sigma_p=1.4(1)k$. The sample is split between the $2J+1$ magnetic sub-levels by a magnetic field gradient. For each magnetic projection, the density profile, taken after time-of-flight, provides the velocity distribution in the $xz$ plane. 
\label{fig_image}}
\end{figure}

\medskip
\titlemethods{Atom imaging and data analysis} 
At the end of the experimental sequence, the light beams are switched off and the atoms are let to expand for a duration of $\SI{2.4}{\milli\second}$ before an absorption image is taken.  During expansion, a magnetic field gradient is applied, leading to a spatial separation of the different magnetic sub-levels $m$ along $z$.  The measured atom density thus gives access to the velocity distribution $\mathcal{D}_{\pbold}(m,\vbold)$, resolved in magnetic projection $m$, for the prepared momentum state $\pbold$. An example of an absorption image is shown in \fig{fig_image}.

\begin{figure*}[!t]
 \includegraphics[
 draft=false,scale=0.99,
 trim={1mm 2mm 0 0.cm},
 ]{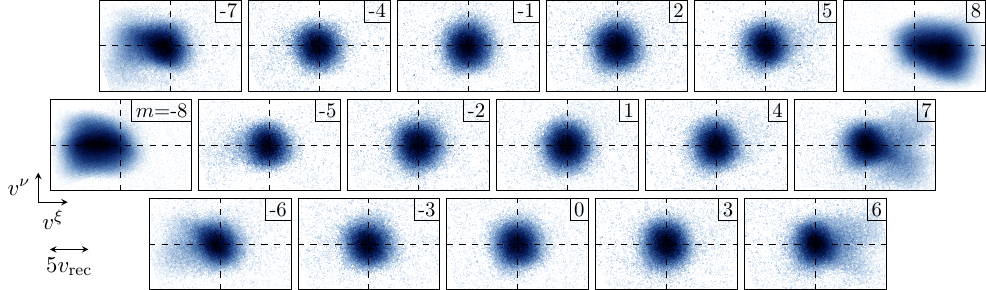}
\caption{
\textbf{Density of states.} Density of states $\mathcal{D}(m,\vbold)$ of the ground band as a function of the magnetic projection $m$ and the velocity $\vbold$. Each panel corresponds to a given $m$ component. The density of states is obtained by integrating the spin-resolved velocity distributions over the momentum interval $|p|\leq12k$. The dashed lines indicate  $v^\xi=0$ and $v^\nu=0$. This data is used for the generation of Fig.\,2 and 3.
\label{fig_DOS}}
\end{figure*}

The thermal momentum width $\sigma_p=1.4(1)k$ after loading in the ground band ($\sigma_p=0.9(1)k$ prior to loading) leads to an averaging of the measured spin-resolved velocity distribution around a mean value $\langle\pbold\rangle$. However, a given momentum state expands on a discrete set of velocities as $M\vbold=\pbold-2km\unitx+\ell \Kbold$, with $\ell$ integer. Different momentum states thus contribute to separate components of the velocity distribution, which allows us to deconvolve the thermal broadening and reconstruct the  velocity distribution for each momentum value. In order to treat all momenta on equal footing, we sum the spin-resolved velocity distributions measured for a dense sample of momenta $\langle\pbold\rangle$ within the Brillouin zone, before performing the momentum deconvolution. Averaging all momenta before deconvolution also suppresses the effect of residual forces, such as gravity and laser beam confinement, occurring during the state preparation. The integrated distribution $\mathcal{D}(m,\vbold)=\int\dd\pbold\,\mathcal{D}_{\pbold}(m,\vbold)$, plotted in \fig{fig_DOS}, physically corresponds to the  spin-velocity density of states of the ground band.

The spin-resolved velocity distributions $\mathcal{D}_{\pbold}(m,\vbold)$ give direct access to the spin projection probabilities $\Pi_m$ and mean velocities $\langle\vbold\rangle$, plotted in \fig{fig_spin_pumping}.
The mean spin projection is then given by
\[
\langle m\rangle\equiv\sum_m\Pi_m m. 
\]
The mean value of the cyclic variable $n$ is a priori ill-defined \cite{lynch_quantum_1995}. However, since we observe a peaked probability distribution in $n$, we track the dynamics along $n$ from the following estimate of the mean position 
\[
\langle n\rangle\equiv \frac{2}{3\pi}\arg\left<\exp\left(\I\frac{2\pi}{3}m\right)\right>.
\]

\medskip
\titlemethods{Response to a force along $m$} 
The measurement of the bulk Berry curvature presented in the main text relies on the application of a force in the $xz$ plane, leading to a Hall drift along $n$ and $m$. In order to access the Berry curvature component $\Omega^{\text{bulk}}_{nm}$, we consider a force applied along $m$, which is equivalent to a perturbative Zeeman field $\hbar\Delta J_z$. This field can be produced by additional detunings of the Raman transitions $\delta_x=-\Delta$ and $\delta_z=2\Delta$. The gauge transform $U=\exp(\I\Delta t J_z)$ suppresses the time-dependency while adding the desired Zeeman field $\hbar\Delta J_z$. As explained earlier, shifts in detuning can also be interpreted as a shift in momentum $\pbold$. A perturbative Zeeman energy is thus equivalent to a variation of momentum within the ground band. Since all momentum states of the ground band are characterized by static spin distributions, we conclude the absence of velocity along $n$ induced by the force along $m$, hence $\Omega^{\text{bulk}}_{nm}=0$.

\medskip
\titlemethods{Berry curvature measurements} 
The second Chern character plotted in \fig{fig_precession}a is obtained as a non-linear combination of Berry curvatures. In this section we show that the latter can be expressed in terms of  $\pbold$-variations of the measured group velocity $\langle\vbold\rangle(\pbold)$ (\fig{fig_velocity}a). 

\begin{figure}[!b]
 \includegraphics[
 draft=false,scale=0.9,
 trim={2mm 2mm 0 0.cm},
 ]{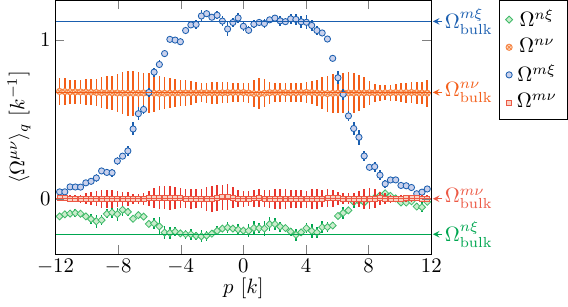}
\caption{
 \textbf{Berry curvature.} 
 Berry curvature components $\Omega^{\alpha\beta}$ computed from the measurements of the mean velocity. 
 The measured values are independent on the quasi-momentum $q$ within error bars, and we show the
$q$-averaged value $\langle\Omega^{\alpha\beta}\rangle_q$ as a function of the momentum $p$. The solid line are the components of the bulk Berry curvature $\boldsymbol\Omega_{\text{bulk}}$ given by \eqP{eq_Omega_bulk}.
\label{fig_Berry_curvature}}
\end{figure}

We consider  an atom initially prepared in a wavepacket of the ground band, centered on a momentum state $\pbold$. Following Laughlin'argument of a charge pump \cite{laughlin_quantized_1981}, the Berry curvature $\Omega^{\alpha\beta}(\pbold)$ quantifies the linear response in position $\delta \langle r^\alpha\rangle$ induced by an adiabatic increase of a gauge field $\delta  A_\beta$, as
 \[
\Omega^{\alpha\beta}=-\frac{\delta\langle r^\alpha\rangle}{\delta  A_\beta}.
\]
An equivalent relation links time-dependent gauge fields to the response in velocity $\delta \langle v^\alpha\rangle$, which is directly accessible in our experiments. More precisely, considering gauge field variations linear in time, as  $\delta  A_\beta=\delta\Gamma_\beta t$, we express the Berry curvature as 
 \begin{equation}\label{eq:Omega}
\Omega^{\alpha\beta}=-\frac{\delta\langle v^\alpha\rangle}{\delta\Gamma_\beta}.
\end{equation}
We now focus on gauge field variations along the synthetic dimensions $m$ and $n$, which induces velocities along spatial dimensions $x$ and $z$. The perturbative gauge fields correspond to additional Peierls phases in the two-photon Raman processes, as
\begin{align*}
\phi_x&=-2kx+(\delta\Gamma_n+\delta\Gamma_m) t,\\
 \phi_z&=-2kz+(\delta\Gamma_n-2\delta\Gamma_m)  t.
\end{align*}
The time dependency can be eliminated by considering the atom in a reference frame moving at velocity 
\begin{align*}
\delta\ubold&=\frac{\delta\Gamma_n(\unitx+\unitz)+\delta\Gamma_m(\unitx-2\unitz)}{2k}\\&=\frac{\delta\Gamma_n}{2\sqrt 5 k}(-\unitxi+3\unitnu)+\frac{\sqrt 5 \delta\Gamma_m}{2k}\unitxi.
\end{align*}
In that frame, the atom occupies the state of momentum $\pbold-M\delta\ubold$ and velocity $\langle  \vbold\rangle(\pbold-M\delta\ubold)$. Coming back to the original reference frame, the change of mean velocity due to the gauge field variation writes
\begin{align}
\delta\langle \vbold\rangle&=\langle  \vbold\rangle(\pbold-M\delta\ubold)+\delta\ubold\nonumber\\
&=\frac{\delta\Gamma_n}{2\sqrt 5 k}\left(-\unitxi+3\unitnu+M\frac{\partial \langle  \vbold\rangle}{\partial p}-3M\frac{\partial \langle  \vbold\rangle}{\partial q}\right)\nonumber\\
&\;\;\;\;+\frac{\sqrt 5 \delta\Gamma_m}{2k}\left(\unitxi-M\frac{\partial \langle  \vbold\rangle}{\partial p}\right).\label{eq:deltav}
\end{align}
This provides a linear relation between velocities along $x$ and $m$ and perturbative gauge fields along $m$ and $n$, from which the Berry curvature components can be deduced according to \eqP{eq:Omega}.
Taking the $\xi$- and $\nu$-projections of \eqP{eq:deltav} yields the Berry curvature components
\begin{align*}
\Omega^{\xi n}&=\frac{1}{2\sqrt 5 k}\left(1-M\frac{\partial\langle v^\xi\rangle}{\partial p}+3M\frac{\partial\langle v^\xi\rangle}{\partial q}\right),\\
\Omega^{\nu n}&=\frac{1}{2\sqrt 5 k}\left(-3-M\frac{\partial\langle v^\nu\rangle}{\partial p}+3M\frac{\partial\langle v^\nu\rangle}{\partial q}\right),\\
\Omega^{\xi m}&=\frac{\sqrt 5}{2k}\left(-1+M\frac{\partial\langle v^\xi\rangle}{\partial p}\right),\\
\Omega^{\nu m}&=\frac{\sqrt 5}{2k}M\frac{\partial\langle v^\nu\rangle}{\partial p}.
\end{align*}
We use these expressions to compute  the Berry curvature tensor from the measurements of the mean velocity $\langle\vbold\rangle(\pbold)$. Our data, shown in \fig{fig_Berry_curvature}, is consistent with the expected Berry curvature $\Omega_{\text{bulk}}$ in the bulk mode region of the Brillouin zone.

\medskip
\titlemethods{Effective continuous model in the bulk} 
\subtitlemethods{Derivation of the continuous model} 
The cyclotron dynamics in the bulk of our system can be modeled using a simple effective model of a charged particles evolving in 4D and subjected to a magnetic field. In the Hamiltonian \eqP{eq_H}, we replace spin ladder operators by translation operators along $m$ and $n$, expressed in terms of momentum operators $p_m$ and $p_n$ as
\begin{align*}
J_+&=\exp[-\I (p_m+p_n)]\sqrt{J(J+1)-m(m+1)} \\
&\simeq(J-m^2/2J)\exp[-\I (p_n+p_m)],
\end{align*}
where we assume  $|m|\ll J$, and  only kept the lowest order in $1/J$. Similarly, we obtain for the second-order spin transition
\[
J_-^2\simeq(J^2-m^2)\exp[-\I (p_n-2p_m)].
\]
The Hamiltonian then writes
\begin{align*}
H=&\frac{Mv^2}{2}-\bigg(\ta\E^{\I \phia}\frac{J_+}{J}+\tb\E^{\I \phib}\frac{J_-^2}{J^2}+\hc\bigg)+\beta\frac{J_z^2}{J^2}\\
\simeq&\,\frac{p_x^2+p_z^2}{2M}-\bigg[t_x\left(1-\frac{1}{2}\left(\frac{m}{J}\right)^2\right)\E^{-\I(p_n+p_m-2kx)}\\
&+t_z\left(1-\left(\frac{m}{J}\right)^2\right)\E^{-\I(p_n-2p_m-2kz)}+\hc\bigg]+\beta\left(\frac{m}{J}\right)^2.
\end{align*}
Expanding it up to quadratic order in all dynamical variables, we obtain
\begin{align}
H&\simeq\epsilon_0+H_{\text{kin}}+\beta'\left(\frac{m}{J}\right)^2,\nonumber\\
\epsilon_0&=-2t_x-2t_z,\nonumber\\
H_{\text{kin}}&=\frac{p_x^2+p_z^2}{2M}+t_x(p_n+p_m-2kx)^2\nonumber\\
&\phantom{=\;} +t_z(p_n-2p_m-2kz)^2,\label{eq_Hkin}\\
\beta'&=\beta+t_x+2t_z.\nonumber
\end{align}
The energy $\epsilon_0$ is a mere energy shift playing no role in the atom dynamics. Reminding that $\beta=-2t_z$, we are left with a residual Zeeman shift $\beta'=t_x$ that explains the residual positive energy curvature in the bulk of the dispersion relation (\fig{fig_scheme}d).

\medskip
\subtitlemethods{Generalized Landau levels} 
The kinetic energy can be recast as
\begin{equation}
H_{\text{kin}}=\frac{1}{2}\left(\frac{1}{M}\right)^{\alpha\beta}(p_\alpha-A_\alpha)(p_\beta-A_\beta),\label{eq_Hkin2}
\end{equation}
where the vector potential
\[
\Abold=\frac{-2k}{3}(0,0,2x+z,x-z)
\]
and its associated magnetic field
 \[
  \Bbold=\frac{2k}{3}\begin{pmatrix}
     0 & 0 & -2 & -1\\
     0 & 0 & -1 & 1\\
     2 & 1 & 0 & 0\\
     1 & -1 & 0 & 0
    \end{pmatrix}
\]
were already introduced in the main text. 
The inverse mass tensor is given by
\[
\frac{1}{\Mbold}=\begin{pmatrix}
1/M & 0 & 0 & 0 \\
0 & 1/M & 0 & 0 \\
0 & 0 & 2t_x+2t_z & 2t_x-4 t_z\\
0 & 0 & 2t_x-4t_z & 2t_x+8 t_z
\end{pmatrix}.
\]

Considered alone, the kinetic Hamiltonian $H_{\text{kin}}$ describes the motion of a charged particle of anisotropic mass and subjected to a magnetic field in 4D. The momenta $p_m$ and $p_n$ are conserved quantities and the $x$ and $z$ motions map to independent harmonic oscillators of frequencies
\begin{equation}
\omega_x=4\sqrt{\Erec t_x},\quad \omega_z=4\sqrt{\Erec t_z}.\label{eq_frequencies_continuous_model}
\end{equation}
The harmonic energy spectrum $E_{n_x,n_z}=\omega_x(n_x+\tfrac{1}{2})+\omega_z(n_z+\tfrac{1}{2})$ corresponds to the direct sum of Landau levels with macroscopic degeneracy indexed by $p_m$ and $p_n$. For the couplings $t_x=5.69(6)\,\Erec$ and $t_z=5.1(1)\,\Erec$  used in our experiments, the expressions \eqP{eq_frequencies_continuous_model} yield $\omega_x=2\pi\times\SI{30.2(2)}{\kilo\hertz}$ and $\omega_z=2\pi\times\SI{28.4(3)}{\kilo\hertz}$, close to the frequencies computed from the actual Hamiltonian \eqP{eq_H}, namely  $\omega_x=2\pi\times\SI{27.4(2)}{\kilo\hertz}$ and $\omega_z=2\pi\times\SI{26.5(3)}{\kilo\hertz}$.

\medskip
\subtitlemethods{Band dispersion due to the quadratic Zeeman shift} 
When including the quadratic Zeeman shift $\beta'(m/J)^2$, the momentum $p_m$ is no longer conserved. To retrieve the energy spectrum in a simple manner, we perform a gauge transform of unitary
\[
U=\exp\left(\frac{2k}{3}\left[(2x+z)n+(x-z)m\right]\right),
\]
leading to a new Hamiltonian
\begin{align}
H&=\epsilon_0+\;\frac{1}{2M}\left(p_x-\frac{2k}{3}(2n+m)\right)^2\nonumber\\
&+\frac{1}{2M}\left(p_z-\frac{2k}{3}(n-m)\right)^2\nonumber\\
&+t_x(p_n+p_m)^2 +t_z(p_n-2p_m)^2+\beta'\left(\frac{m}{J}\right)^2\label{eq_Hkin}.
\end{align}
This Hamiltonian conserves the  momenta $p_x$ and $p_z$. The $m$ and $n$ dynamics is quadratic, such as it is straightforward to compute the energy spectrum, as 
\[
E_{n_x,n_z}(p)=\epsilon_0+\omega_x(n_x+\tfrac{1}{2})+\omega_z(n_z+\tfrac{1}{2})+\frac{5\beta'}{4J^2} \left(\frac{p}{k}\right)^2.
\]
where $p$ is the momentum component along $\unitxi$ and we assumed $\beta'\ll J^2\Erec$ for simplicity. This expression explains the $p$ curvature of the ground band shown in \fig{fig_scheme}d, while the bulk spectrum is essentially flat along $q$.

\medskip
\subtitlemethods{Equations of motion} 
The continuous model can also be used to compute the cyclotron dynamics. For simplicity, we restrict the dynamics to  the kinetic Hamiltonian \eqP{eq_Hkin}, and obtain the equations of motion
\[
\partial_t v^\alpha=\left(\frac{1}{M}\right)^{\alpha\beta}B_{\beta\gamma}v^\gamma,
\]
explicitely given by
\begin{equation*}
\partial_t
\begin{pmatrix}
 v_x \\
 v_z \\
 v_n\\
 v_m
\end{pmatrix}
\!
=
\!
\begin{pmatrix}
0&0&-4\vrec/3&-2\vrec/3\\
0&0&-2\vrec/3&2\vrec/3\\
4k t_x&4k t_z&0&0\\
4k t_x&-8k t_z&0&0\\
\end{pmatrix}
\!
\begin{pmatrix}
 v_x \\
 v_z \\
 v_n\\
 v_m
\end{pmatrix}
\!. \label{eq_dynamics}
\end{equation*}
The dynamics in the $xz$ plane can be expressed in a closed manner  as
\begin{align*}
\partial_t^2\begin{pmatrix}
 v_x \\
 v_z 
\end{pmatrix}&=
\begin{pmatrix}
-4\vrec/3&-2\vrec/3\\
-2\vrec/3&2\vrec/3
\end{pmatrix}
\begin{pmatrix}
4k t_x&4k t_z\\
4k t_x&-8k t_z
\end{pmatrix}
\begin{pmatrix}
 v_x \\
 v_z 
\end{pmatrix}\\
&=
-\begin{pmatrix}
\omega_x^2&0\\
0&\omega_z^2
\end{pmatrix}
\begin{pmatrix}
 v_x \\
 v_z 
\end{pmatrix},
\end{align*}
describing harmonic motion along $x$ and $z$ of frequencies $\omega_x$ and $\omega_z$ given by \eqP{eq_frequencies_continuous_model}. 

\begin{figure*}[]
 \includegraphics[
 draft=false,scale=0.99,
 trim={2mm 2mm 0 0.cm},
 ]{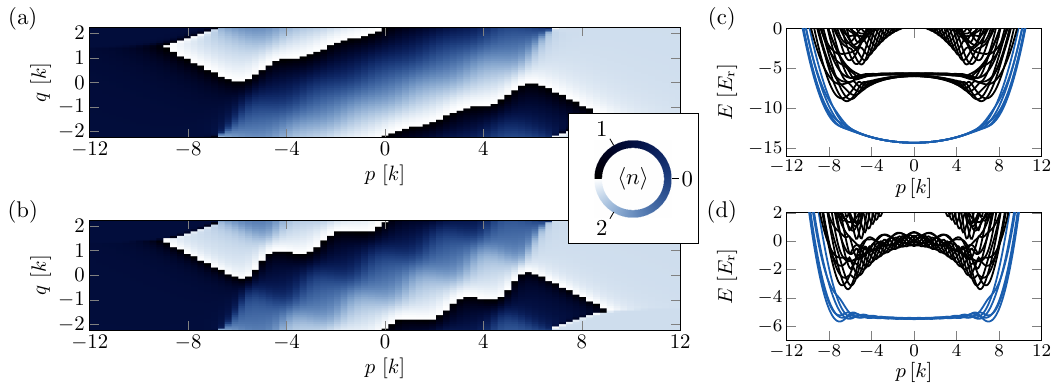}
\caption{
 \textbf{Validity regime of the continuous model.}
(a) Evolution of the mean value $\langle n\rangle$ as a function of momentum throughout the ground band, calculated for the parameters $t_x=5.7\,\Erec$ and  $t_z=5.1\,\Erec$ as used in our experiments. (b) Same for smaller couplings $t_x=2.85\,\Erec$ and  $t_z=2.55\,\Erec$. (c,d) Dispersion relations corresponding to the couplings used in (a,b). 
\label{fig_continuous_validity}}
\end{figure*}

\medskip
\subtitlemethods{Domain of validity of the continuous model} 

The continuous model provides an accurate description of our system as long as the discrete nature of the synthetic dimensions can be neglected. This is the case when the ground-band orbitals are sufficiently delocalized in spin space. For the laser couplings used in our experiments,  the characteristic rms widths of orbitals, $\sigma_m=1.9(1)$ and $\sigma_n=1.2(1)$ along $m$ and $n$, exceed unity, such that the discreteness of the synthetic dimension has only a weak effect. Thus, we expect the continuous model to be valid to a good approximation, which implies that the system can be expressed as a direct sum of a pair of 2D Landau levels. In particular, we expect the mean spin projections $\langle m\rangle$ and $\langle n\rangle$ to be coupled to a single spatial direction, along  $2\unitx+\unitz$ and $\unitx+\unitz$, respectively (see  \fig{fig_continuous_validity}a).

This simple behavior breaks down for smaller light couplings due to the reduction of the orbital width in spin space. As shown in  \fig{fig_continuous_validity}b, for coupling strengths $t_x=2.85\,\Erec$ and $t_z=2.55\,\Erec$ (half of the values used in our experiments), the mean spin projection $\langle n\rangle$ shows a more complex dependency on the momentum $\pbold$. This shows that the system can no longer be decomposed  as the direct sum of two Landau levels. Nevertheless, the ground band remains gapped in the bulk (\fig{fig_continuous_validity}d), such that the non-trivial topology persists with  a local Chern marker $C_2=1$.

\begin{figure}[]
 \includegraphics[
 draft=false,scale=0.99,
 trim={2mm 2mm 0 0.cm},
 ]{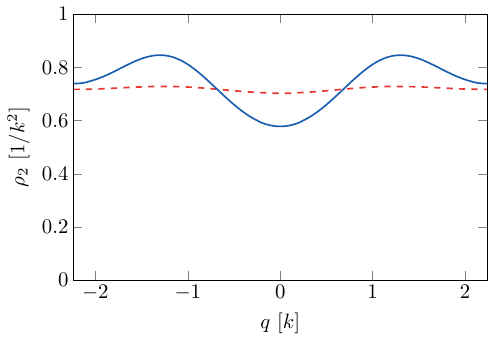}
\caption{
Variation of the second Chern character $\rho_2$ with $q$, for a fixed $p=0$, and for two different couplings  $t_x=5.7\,\Erec$ and  $t_z=5.1\,\Erec$ (red dashed line) and  $t_x=2.85\,\Erec$ and  $t_z=2.55\,\Erec$ (blue solid line). 
\label{fig_rho2}}
\end{figure}

Another illustration of the validity of the continuous model is shown in \fig{fig_rho2}, which shows the variation of the second Chern character along a line $p=0$, $q$ arbitrary, for the two coupling strengths considered above. We find that $\rho_2$ is almost constant for the largest coupling strength, in agreement with the continuous model, while it is strongly modulated for the smallest couplings.

\newpage

\medskip
\titlemethods{Foucault pendulum precession from a  $b_{nm}$ field} 
We investigate the effect of a magnetic perturbation $b_{nm}$ in the $nm$ plane within the frame of the continuous model described above.
The  $b_{nm}$ field is induced by an additional phase factor in the $x$-Raman transitions $\phi_x'=-\tfrac{3}{2}b_{nm}m$, such that the vector potential becomes
\[
\Abold=\left(\!0,0,\frac{-2k}{3}(2x+z)-b_{nm}m,\frac{-2k}{3}(x-z)-\frac{1}{2}b_{nm}m\!\right)\!,
\]
yielding a magnetic field
 \[
  \Bbold=\begin{pmatrix}
     0 & 0 & -\frac{4k}{3} & -\frac{2k}{3}\\
     0 & 0 & -\frac{2k}{3} & \frac{2k}{3}\\
     \frac{4k}{3} & \frac{2k}{3} & 0 & b_{nm}\\
     \frac{2k}{3} & -\frac{2k}{3} & -b_{nm} & 0
    \end{pmatrix}
\]
and equations of motion
\begin{align*}
&\partial_t
\begin{pmatrix}
 v_x \\
 v_z \\
 v_n\\
 v_m
\end{pmatrix}
=\\
&\!\begin{pmatrix}
0&0&-4\vrec/3&-2\vrec/3\\
0&0&-2\vrec/3&2\vrec/3\\
4k t_x&4k t_z&-2b_{nm} (t_x-2t_z)&-2b_{nm} (t_x+t_z)\\
4k t_x&-8k t_z&2b_{nm} (t_x+4t_z)&2b_{nm} (t_x-2t_z)\\
\end{pmatrix}
\!
\!
\!
\begin{pmatrix}
 v_x \\
 v_z \\
 v_n\\
 v_m
\end{pmatrix}
\!\!. \label{eq_dynamics_with_bmr}
\end{align*}
In order to obtain a closed equation for the dynamics in the $xz$ plane, we invert the relation
\[
\partial_t
\begin{pmatrix}
 v_x\\
 v_z
\end{pmatrix}
=
\frac{2\vrec}{3}
\begin{pmatrix}
-2&-1\\
-1&1
\end{pmatrix}
\begin{pmatrix}
 v_n\\
 v_m
\end{pmatrix}
\]
as
\[
\begin{pmatrix}
 v_n\\
 v_m
\end{pmatrix}
=
\frac{1}{2\vrec}
\begin{pmatrix}
-1&-1\\
-1&2
\end{pmatrix}
\partial_t
\begin{pmatrix}
 v_x\\
 v_z
\end{pmatrix}.
\]
This allows us to eliminate the $m$ and $n$ velocities and obtain closed dynamics equations along $x$ and $z$, as
\begin{align*}
\partial_t^2 v_x&=-\omega_x^2(v_x+M\rho_2^{\text{bulk}}\partial_t v_z),\\
\partial_t^2 v_z&=-\omega_z^2(v_z-M\rho_2^{\text{bulk}}\partial_t v_x),
\end{align*}
involving the bulk second Chern character $\rho_2^{\text{bulk}}=3/4k^2$.

The Foucault pendulum precession occurs for equal frequencies $\omega_x=\omega_z=\omega$, leading to an $xz$ dynamics
\[
\partial_t^2\vbold=-\omega^2\vbold+2\omega_{\text p}\unitn\times\vbold,
\]
$\unitn=\unitx\times\unitz$  is normal to the $xz$ plane. We recognize the equation of motion of a Foucault pendulum of harmonic frequency $\omega$ and precession rate $\omega_{\text p}$ given by \eqP{eq_omegap} in the main text.

This connection between the precession rate $\omega_{\text p}$ and the second Chern character $\rho_2^{\text{bulk}}$ is not specific to the precise algebra of our setting, and would apply for arbitrary implementations of a 4D quantum Hall system using an artificial magnetic field.

The non-linear electromagnetic response could in principle be measured more directly, from the geometrical drift induced by both a force $f_x$ along $x$ and a magnetic perturbation $b_{nm}$, as
\begin{equation}\label{eq_non_linear_drift}
 \langle z\rangle_{\text{non-linear}}(t)=-\rho_2 b_{nm}f_x t.
\end{equation}
For the $b_{nm}$ field accessible in our setting, we expect a displacement $\delta z\simeq\SI{5}{\nano\meter}$ for an experiment duration $t=k/f_x$  such that the force $f_x$ changes the momentum by $k$. This displacement is too small to be accessible in our setting. We note that the $b_{nm}$ field used in our experiments is already at the limit of validity of the linear response given by \eqP{eq_non_linear_drift}.

\medskip
\titlemethods{Local second Chern marker}
The occurrence of edge states in our system implies the absence of a gap in the energy spectrum, such that the system as a whole is not topological (\fig{fig_scheme}d). Nevertheless, the lowest band features a weak dispersion and a bulk energy gap, such that non-trivial topological properties are expected when probing the bulk only. To access it, we consider the local second Chern marker, defined as  \cite{bianco_mapping_2011,sykes_local_2021} 
\[
C_2(\rbold)=2\pi^2\epsilon_{\alpha\beta\delta\gamma}\langle\rbold|Pr^\alpha P r^\beta P r^\delta P r^\gamma P|\rbold\rangle,
\]
where $P$ projects on the lowest Bloch band. For an infinite and homogeneous system, it  coincides with the integer second Chern number $\mathcal{C}_2$. In a finite system with open boundaries, we expect $C_2(\rbold)$ to match  $\mathcal{C}_2$ for positions $\rbold$ within the bulk only \cite{bianco_mapping_2011}.

\begin{figure}
 \includegraphics[
 draft=false,scale=0.85,
 trim={1mm 2mm 0 0.cm},
 ]{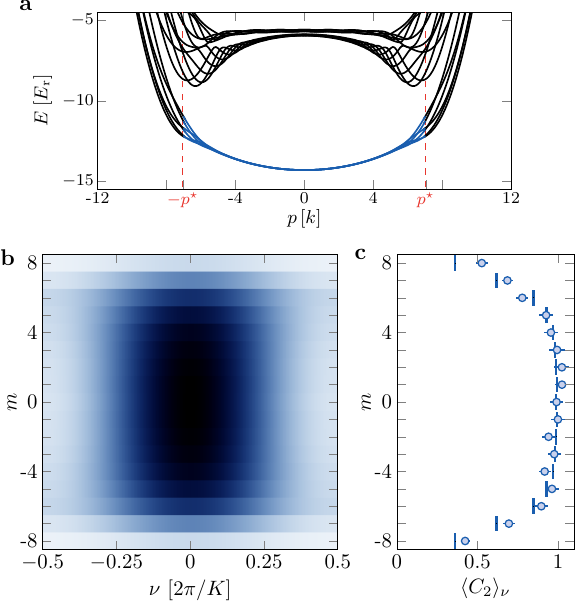}
\caption{
\textbf{Local second Chern marker.} \textbf{a.} Dispersion relation computed for the parameters used in the experiment. The calculation of $C_2(\nu,m)$ involves all states of the ground band in the interval $(-p^\star,p^\star)$, with $p^\star=7k$, discarding strongly dispersive edge modes. 
\textbf{b.} Local second Chern marker $C_2$ plotted as a function of $\nu$ and $m$.
\textbf{c.} $\nu$-average of $C_2$ as a function of $m$,  compared to the experimental data (dots).
\label{fig_LSCM}}
\end{figure}

For our system, this marker can be rewritten in terms of accessible observables as follows \cite{chalopin_probing_2020}. We decompose the band projector $P=\int \dd p\,\dd q\ket{\psi_{pq}}\bra{\psi_{pq}}$ on magnetic Bloch states $\ket{\psi_{pq}}$, such that the Chern marker writes
\begin{align}
C_2(\rbold)&=2\pi^2\epsilon_{\alpha\beta\delta\gamma}\int \dd p_1\,\dd q_1\,\dd p_2\,\dd q_2\,\dd p_3\,\dd q_3\nonumber\\
&\;\;\;\;\;\;\;\;\;\;\psi_{p_1q_1}^*(\rbold)\psi_{p_2q_2}(\rbold)
c^{\alpha\beta}_{p_1q_1p_3q_3}c^{\delta\gamma}_{p_3q_3p_2q_2},\label{eq_SLCM}\\
c^{\alpha\beta}_{p_1q_1p_2q_2}& \equiv\langle\psi_{p_1,q_1}|r^\alpha P r^\beta|\psi_{p_2,q_2}\rangle.\nonumber
\end{align}
The coefficient involving $\xi$ and $m$ directions reads
\[
c^{\xi m}_{p_1q_1p_2q_2}=\I\langle m\rangle_{p_2q_2}\langle\psi_{p_1q_1}|\psi_{p_2q_2}\rangle \delta'(p_1-p_2)\delta(q_1-q_2),
\]
and similar expressions can be written for other coefficients $c^{\alpha\beta}_{p_1q_1p_2q_2}$. Injecting these expressions in \eqP{eq_SLCM}, we obtain
\begin{align*}
C_2(\rbold)=4\pi^2\!\!\int  \!\dd p\,\dd q\, |\psi_{pq}(\rbold)|^2(\partial_p \langle m \rangle\partial_q \langle n\rangle-\partial_q \langle m \rangle\partial_p \langle n \rangle).
\end{align*}
We recognize the Berry curvature components, e.g. $\Omega^{m\xi}=\partial_p \langle m \rangle$, such that their non-linear combination matches the second Chern character $\rho_2$, leading to
\begin{align*}
C_2(\rbold)=4\pi^2\int  \dd p\,\dd q\, |\psi_{pq}(\rbold)|^2\rho_2(p,q).
\end{align*}
The system is transitionally invariant along $\xi$, and exhibits a discrete translation symmetry along $\nu$ of reciprocal lattice vector $\Kbold$. We thus expect the local second Chern character to depend on $\nu$ and $m$ only. 
Since the $\nu$ modulation occurs on a small length scale $1/K$ only, we define the $\nu$ coarse average  $\langle C_2\rangle_\nu(m)$, involving the $m$-projection probability $\Pi_m$ only, as 
\[
\langle C_2\rangle_\nu(m)=\frac{1}{3}\int\dd p\,\dd q\, \Pi_m(p,q)\rho_2(p,q).
\]
The integration should be performed on bulk modes only, defined by the interval $|p|<p^*$. The momentum cutoff $p^*=7k$ is chosen such that the ground band energy approximately reaches at $p^*$ the middle of the bulk gap between the ground and first excited bands (\fig{fig_LSCM}a). The precise choice of $p^\star$ does not significantly influence the values of $\langle C_2\rangle_\nu(m)$ in the bulk of the system.

We show in \fig{fig_LSCM}b the local second Chern marker $C_2(\nu,m)$, numerically computed for the coupling parameters used in the experiment. Its $\nu$-average $\langle C_2\rangle_\nu(m)$ is very close to the second Chern number $\mathcal{C}_2=1$ in the bulk $|m|\lesssim 5$, and agrees well with our measurements (\fig{fig_LSCM}c).

\end{document}